\documentclass[aps,prb,twocolumn,amsmath]{revtex4}
\usepackage{latexsym}
\usepackage{amssymb}
\usepackage{graphicx}
\usepackage{bm}

\newcommand{\pdag}{{\phantom{\dagger}}}
\newcommand{\bq}{\begin{equation}}
\newcommand{\eq}{\end{equation}}
\newcommand{\bn}{\begin{eqnarray}}
\newcommand{\en}{\end{eqnarray}}

\begin{document}

\title{Shot noise in resonant tunneling through an interacting quantum dot with intradot 
spin-flip scattering\footnote{Based on work presented at the 2004
IEEE NTC Quantum Device Technology Workshop}}
\author{Ivana Djuric$^{1}$, Bing Dong$^{1,2}$,\cite{db} and H. L. Cui$^{1,3}$}
\affiliation{$^{1}$Department of Physics and Engineering Physics, Stevens Institute of 
Technology, Hoboken, New Jersey 07030 \\
$^{2}$Department of Physics, Shanghai Jiaotong University,
1954 Huashan Road, Shanghai 200030, China \\
$^{3}$School of Optoelectronics Information Science and Technology, Yantai University, 
Yantai, Shandong, China}

\begin{abstract}

In this paper, we present theoretical investigation of the zero-frequency shot noise 
spectra in electron tunneling through an interacting quantum dot connected to two 
ferromagnetic leads with possibility of spin-flip scattering between the two spin states 
by means of the recently developed bias-voltage and temperature dependent quantum rate 
equations. For this purpose, a generalization of the traditional 
generation-recombination approach is made for properly taking into account the coherent 
superposition of electronic states, i.e., the nondiagonal density matrix elements. Our 
numerical calculations find that the Fano factor increases with increasing the 
polarization of the two leads, but decreases with increasing the intradot spin-flip 
scattering.

\end{abstract}

\maketitle

\section{INTRODUCTION}

Recent rapid development of spintronic and single electron devices
has resulted in intensive studies of spin-related
phenomena in various resonant tunneling structures.\cite{Prinz,Wolf}
One of the devices of interest is a quantum dot (QD), sandwiched
between two ferromagnetic electrodes.\cite{dot1,dot2,dot3} This kind of spin-related 
single-electron devices suffers inevitably intrinsic relaxations (decoherence) due to 
the spin-orbital interaction\cite{Fedichkin} or the hyperfine-mediated spin-flip 
transition.\cite{Erlingsson} The effect of the intrinsic decoherence on the quantum 
transport properties in these devices still deserves more investigation. As is well 
known, measurement of shot noise, which is a consequence of the discrete nature of the 
electric charge, can provide information about the microscopic interactions and the 
statistics of particles, which is not available through the conductance measurement 
alone.\cite{Beenakker,Blanter} Therefore, it is remarkably desirable to expand our 
knowledge about the shot noise spectra in such a spin-related single-electron tunneling 
device.
 
For metallic QD without intrinsic decoherence, the zero-frequency Fano factor, which 
quantifies correlations with respect to the uncorrelated Poissonian noise, was analyzed 
with the classical rate equation approach (classical shot noise) in 
Refs.[\onlinecite{Hershfield,Korotkov,Hanke}]. However, in order to account for the 
quantum effects during sequential tunneling processes, such as coherence and 
superposition of the wave functions and various intrinsic interactions, a more elaborate 
quantum description is required. The quantum rate equation approach is a suitable tool 
for accomplishing this task.\cite{Nazarov,Gurvitz,Fedichkin,Dong} In particular, some of 
the authors, Dong, Cui, and Lei, recently established the quantum rate equations for 
coherent tunneling through the coupled quantum systems with a bias-voltage and 
temperature dependent version.\cite{Dong} Consequently, these equations allow us to 
study the quantum tunneling in moderately small bias-voltage region, where the quantum 
coherence plays more prominent role. Therefore, the purpose of the present paper is to 
develop an approach for calculation of the coherent effect, due to the intrinsic spin 
scattering, on the zero-frequency shot noise spectra in an interacting QD weakly 
connected with two ferromagnetic leads based on the quantum rate equations.

The rest of this paper is organized as follows. In section II, we introduce the 
Hamiltonian for the interacting QD with intradot spin scattering, and then give the 
quantum rate equations. In section III, we describe our methodology for evaluating the 
quantum shot noise. In section IV, we present the numerical results and discussions. 
Finally, a brief summary is presented in section V.

\section{SINGLE QUANTUM DOT: Model and quantum rate equations}

The system that we study is a single QD with an arbitrary intradot Coulomb interaction 
$U$ connected with two ferromagnetic leads. In this paper, we assume that the tunneling 
coupling between the dot and the leads is weak enough to guarantee no Kondo effect and 
that the QD is in the Coulomb blockade regime. For simplicity, we model the intrinsic 
spin relaxation with a phenomenological spin-flip term $R$ and assume that this 
spin-flip process just happen in the dot (the spin conserving tunneling). 

Moreover, we assume that the temperature is small enough to see the effects due to the
discrete charging and discrete structure of the energy levels,
i.e. $k_{B}T\ll U$,$\Delta$, where, $\Delta$ is the
energy spacing between orbital levels. Each of the two leads is separately in thermal 
equilibrium with the chemical potential $\mu_{\eta}$, which is assumed to be zero in the 
equilibrium condition and is taken as the energy reference throughout the paper. In the 
nonequilibrium case, the chemical potentials of the leads differ
by the applied bias $V$. We are interested in the regime $eV\ll
\Delta$, where only one dot level $(\epsilon_{d})$ contributes to
the transport. Here we neglect Zeeman splitting of the level due to weak magnetic fields 
$B$ ($g\mu_{B}B<k_{B}T)$, which means that both the spin-up and
spin-down transports through the dot go through the same orbital
level $\epsilon_{d}$. Therefore, the Hamiltonian of resonant tunneling through a single 
QD
can be written as:\cite{Dong}
\bn 
H &=& \sum_{\eta,
k, \sigma}\epsilon _{\eta k\sigma} c_{\eta k\sigma }^{\dagger} c_{\eta k\sigma}^{\pdag}+ 
\epsilon_{d} \sum_{\sigma}
c_{d \sigma }^{\dagger }c_{d \sigma }^{\pdag}+ R( c_{d \uparrow}^{\dagger} 
c_{d\downarrow}^{\pdag} \cr
&& + c_{d\downarrow}^{\dagger} c_{d\uparrow}^{\pdag}) + Un_{d \uparrow }n_{d \downarrow} 
+ \sum_{\eta, k, \sigma} (V_{\eta \sigma} c_{\eta k\sigma }^{\dagger }c_{d 
\sigma}^{\pdag} +{\rm {H.c.}}), \cr && \label{hamiltonian1}
\en
where $c_{\eta k \sigma}^{\dagger}$ ($c_{\eta k \sigma }$) and $c_{d \sigma}^{\dagger}$ 
($c_{d \sigma}$) are the creation (annihilation) operators for electrons with momentum 
$k$, spin-$\sigma$ and energy $\epsilon_{\eta k \sigma}$ in the lead $\eta$ ($={\rm 
L,R}$) and for a spin-$\sigma$ electron on the QD, respectively. $n_{d\sigma}=c_{d 
\sigma}^{\dagger} c_{d \sigma}^{\pdag}$ is the occupation operator in the QD. The fourth 
term describes the Coulomb interaction among electrons on the QD. The fifth term 
represents the tunneling coupling between the QD and the reservoirs. We assume that the 
coupling strength $V_{\eta \sigma}$ is spin-dependent being capable of describing the 
ferromagnetic leads.

Under the assumption of weak coupling between the QD and the leads, and
applying the wide band limit in the two leads, electronic
transport through this system in sequential regime can be described by the bias-voltage 
and temperature dependent quantum rate equations for the dynamical evolution of the 
density matrix elements:\cite{Dong}
\begin{subequations}
\label{rateqSQD} 
\bq
\dot{\rho}_{00}= \sum_{\eta\sigma} ( \Gamma_{\eta\sigma}^{-} \rho_{\sigma \sigma} - 
\Gamma_{\eta\sigma}^{+} \rho_{00}), \label{r0} 
\eq 
\bn
\dot{\rho}_{\sigma \sigma} &=& \sum_{\eta}\Gamma_{\eta\sigma}^{+} 
\rho_{00}+\sum_{\eta}\widetilde {\Gamma}_{\eta\bar{\sigma}}^{-} \rho_{dd}- \sum_{\eta}( 
\Gamma_{\eta\sigma}^{-} + \widetilde {\Gamma}_{\eta\bar{\sigma}}^{+}) \rho_{\sigma 
\sigma}\nonumber \\
&& + i R(\rho_{\sigma \bar{\sigma}}-\rho_{\bar{\sigma}\sigma}), \label{r1} \en \bq
\dot{\rho}_{\sigma \bar{\sigma}} = iR(\rho_{\sigma \sigma} - \rho_{\bar{\sigma} 
\bar{\sigma}})  - \frac{1}{2}\sum_{\eta} ( \widetilde\Gamma_{\eta\sigma}^{+} + 
\widetilde\Gamma_{\eta\bar{\sigma}}^{+} + \Gamma_{\eta\sigma}^{-} + 
\Gamma_{\eta\bar{\sigma}}^{-}) \rho_{\sigma \bar{\sigma}}, \label{r2} \eq \bq
\dot{\rho}_{dd}= \sum_{\eta}\widetilde{\Gamma}_{\eta\downarrow}^{+} \rho_{\uparrow 
\uparrow} + \sum_{\eta}\widetilde {\Gamma}_{\eta\uparrow}^{+} \rho_{\downarrow 
\downarrow} - \sum_{\eta}( \widetilde{\Gamma}_{\eta\uparrow}^{-} + 
\widetilde{\Gamma}_{\eta\downarrow}^{-}) \rho_{dd}, \label{r3} 
\eq
\end{subequations}
($\sigma=\uparrow,\downarrow$ stands for electron spin and $\bar \sigma$ is the spin 
opposite to $\sigma$). The statistical expectations of the diagonal elements of the 
density matrix, $\rho_{ii}$ ($i=\{0, \sigma, d\}$), give the occupation probabilities of 
the resonant level in the QD as follows: $\rho_{00}$ denotes the occupation probability 
that central region is empty, $\rho_{\sigma\sigma}$ means that the QD is singly occupied 
by a spin-$\sigma$ electron, and $\rho_{dd}$ stands for the double occupation by two 
electrons with different spins. Note that they must satisfy the normalization relation 
$\rho_{00}+ \rho_{dd}+ \sum_{\sigma} \rho_{\sigma \sigma}=1$. The non-diagonal elements 
$\rho_{\sigma \bar\sigma}$ describe the coherent superposition of different spin states. 
These temperature-dependent tunneling rates are defined as $\Gamma_{\eta\sigma}^{\pm}= 
\Gamma_{\eta \sigma} f_{\eta}^{\pm}(\epsilon_{d})$ and 
$\widetilde{\Gamma}_{\eta\sigma}^{\pm}= \Gamma_{\eta \sigma} 
f_{\eta}^{\pm}(\epsilon_{d}+U)$, where $\Gamma_{\eta\sigma}$ are the tunneling 
constants, $f_{\eta}^{+}(\omega)=\{1+e^{(\omega-\mu_{\eta})/T} \}^{-1}$ is the Fermi 
distribution function of the $\eta$ lead and 
$f_{\eta}^{-}(\omega)=1-f_{\eta}^{+}(\omega)$. Here, $\Gamma_{\eta\sigma}^{+}$ 
($\Gamma_{\eta\sigma}^{-}$) describes the tunneling rate of electrons with spin-$\sigma$ 
into (out of) the QD from (into) the $\eta$ lead without the occupation of the 
$\bar{\sigma}$ state. Similarly, $\widetilde {\Gamma}_{\eta\sigma}^{+}$ ($\widetilde 
{\Gamma}_{\eta\sigma}^{-}$) describes the tunneling rate of electrons with spin-$\sigma$ 
into (out of) the QD, when the QD is already occupied by an electron with 
spin-$\bar{\sigma}$, revealing the modification of the corresponding rates due to the 
Coulomb repulsion. The particle current $I_{\eta}$ flowing from the lead $\eta$ to
the QD is
\bq
I_{\eta}/e=\sum_{\sigma} ( \widetilde{\Gamma}_{\eta \sigma}^{-} \rho_{dd} + \Gamma_{\eta 
\sigma}^{-} \rho_{\sigma \sigma} - \widetilde{\Gamma}_{\eta \bar{\sigma}}^{+} 
\rho_{\sigma \sigma} - \Gamma_{\eta \sigma}^{+} \rho_{00}).
\label{iii} \eq
This formula demonstrates that the current is totally determined by the diagonal 
elements of the density matrix of the central region. However, the nondiagonal element 
of the density matrix is coupled with the diagonal elements in the rate 
equation~(\ref{r1}), and therefore indirectly influences the tunneling current.

\section{Quantum shot noise formula}

There is a well-established procedure, namely, the generation-recombination approach for 
multielectron channels, for the calculation of the noise power spectrum based on the 
classical rate equations (classical shot noise).\cite{Hershfield,Korotkov,Hanke} In this 
section we modify this approach in order to take into account the nondiagonal density 
matrix elements and derive the general expression for a quantum shot noise for the 
single QD.

Before proceeding with investigation of current correlation, it is helpful to rewrite 
the quantum rate equations as matrix form:
\bq
 \frac{d {\bm \rho}(t)}{dt}= \mathbf{M}\ {\bm \rho}(t)\ ,
 \label{rate}
\eq
where ${\bm \rho}(t)=(\rho_{00}, \rho_{\uparrow \uparrow}, \rho_{\downarrow \downarrow}, 
\rho_{dd}, \rho_{\uparrow \downarrow}, \rho_{\downarrow \uparrow})^{T}$ is a vector 
whose components are the density matrix elements, and the $6 \times 6$ matrix ${\bf M}$ 
can be easily obtained from Eqs.~(\ref{rateqSQD}). Correspondingly, we can write the 
average electrical currents across the left ($I_{L}$) and
right ($I_{R}$) junctions at time $t$ as:
\bq
 \langle I_{L(R)}(t)\rangle=-e \sum_{k} \left[ \hat{\Gamma}_{L(R)} {\bm \rho}(t)
 \right]_{k},
 \label{current}
\eq 
where $\hat{\Gamma}_{L}$ and $\hat{\Gamma}_{R}$ are current
operators and the summation goes over all vector
$[\hat{\Gamma}_{L(R)}{\bm \rho}(t)]$ elements ($k=1,2,\cdots,6$). The current operators
contain the rates for tunneling across the left and right
junctions respectively,\cite{Hershfield} and they can be read from
Eq.~(\ref{iii}) as: 
\bq
\hat{\Gamma}_{\eta}=\pm\left(
\begin{array}{cccccc}
  0 & \Gamma_{\eta\uparrow}^{-} & \Gamma_{\eta\downarrow}^{-} & 0 & 0 & 0 \\
  -\Gamma_{\eta\uparrow}^{+} & 0 & 0 & \tilde{\Gamma}_{\eta\downarrow}^{-} & 0 & 0 \\
  -\Gamma_{\eta\downarrow}^{+} & 0 & 0 & \tilde{\Gamma}_{\eta\uparrow}^{-} & 0 & 0 \\
  0 & -\tilde{\Gamma}_{\eta\downarrow}^{+} & -\tilde{\Gamma}_{\eta\uparrow}^{+} & 0 & 0 
& 0 \\
  0 & 0 & 0 & 0 & 0 & 0 \\
  0 & 0 & 0 & 0 & 0 & 0 \\
\end{array}
\right),
\label{current operators}
\eq
where the sign of the current is chosen to be positive when the
direction of the current is from left to right, so that the $+$
sign in the last equation is for $\eta=L$ and the $-$ sign stands
for $\eta=R$. The stationary current can be obtained as:
\bn
\label{stationary current} 
I = e\sum_{k} \left [\hat{\Gamma}_L {\bm \rho}^{(0)}\right ]_{k} = e \sum_{k} \left [ 
\hat{\Gamma}_R {\bm \rho}^{(0)}\right ]_{k},
\en
where ${\bm \rho}^{(0)}$ is the steady state
solution of Eq.~(\ref{rate}) and which can be obtained from
\bn
\label{steadystate} 
\mathbf{M}{\bm \rho}^{(0)}=0, 
\en
along with the normalization relation $\sum_{n=1}^{4} {\bm \rho}^{(0)}_{n}=1$. We would 
like to point out that in our quantum version of rate equations, it is easy to check 
$\sum_{n} {\bf M}_{nm}=0\ (m=1,2,3,4)$, which implies that: 1) the Matrix ${\bf M}$ has 
a zero eigenvalue; 2) there is always a steady state solution ${\bm \rho}^{(0)}$; 3) the 
normalization relation $\sum_{n=1}^{4} {\bm \rho}_{n}(t)=1$ is independent on time.   

It is well known that the noise power spectra can be expressed as the Fourier transform 
of the current-current correlation function: 
\bn
\label{powerspectrum}
S_{I_{\eta}I_{\eta'}}(\omega)&=&2\int_{-\infty}^{\infty}dt
e^{i \omega t}[\langle
I_{\eta}(t)I_{\eta'}(0)\rangle - \langle I_{\eta} \rangle \langle I_{\eta'} \rangle ] 
\nonumber\\
&=& 2\langle
I_{\eta}(t) I_{\eta'}(0) \rangle_{\omega}- 2\langle I_{\eta}
\rangle_{\omega} \langle I_{\eta'} \rangle_{\omega}.
\en
A convenient way to evaluate the double-time correlation function is to define the 
propagator $\hat{T}(t)=\exp[{\bf M}t]$, which governs the time evolution of the density 
matrix elements $\rho_{k}(t)$. The average value of the
electrical currents across the left ($I_{L}$) and the right
($I_{R}$) junctions at a time $t$ is given by 
\bq
\langle{I_{L(R)}(t)}\rangle=-e \sum_{k} \left[
\hat{\Gamma}_{L(R)}\hat{T}(t) {\bm \rho}^{(0)}\right]_{k},
\label{current1} 
\eq
which allows us to switch the time evolution from the vector ${\bm \rho}(t)$ to the 
current operators. Thus, we identify
$\hat{\Gamma}_{L(R)}(t)=\hat{\Gamma}_{L(R)}\hat{T}(t)$ as
the time-dependent current operators. With these time-dependent
operators we can calculate correlation functions of two current operators
taken at different moments in time. In particular, correlation
function of the currents $I_{\eta}$ and $I_{\eta'}$ in the tunnel
junctions $\eta$ and $\eta'$, measured at the two times $t$ and $0$ respectively, is 
given by\cite{Hershfield}
\bn
\langle{I_{\eta}(t)I_{\eta'}(0) }\rangle
&=& \theta(t)\sum_{k} [\hat{\Gamma}_{\eta}(t)\hat{\Gamma}_{\eta'} {\bm \rho}^{(0)}]_{k} 
\cr 
&+& \theta(-t)\sum_{k}[\hat{\Gamma}_{\eta'}(-t)\hat{\Gamma}_{\eta} {\bm \rho}^{(0)}]_{k} 
\nonumber\\
&=& \theta(t) \sum_{k} [\hat{\Gamma}_{\eta} \hat{T}(t) \hat{\Gamma}_{\eta'} {\bm 
\rho}^{(0)} ]_{k} \cr
&+&  \theta(-t)\sum_{k} [\hat{\Gamma}_{\eta'} \hat{T}(-t) \hat{\Gamma}_{\eta} {\bm 
\rho}^{(0)}]_{k},
\label{current correlation}
\en
where $\theta(t)$ is the Heaviside function and
the two terms in Eq.~(\ref{current correlation}) stand for $t>0$
and for $t<0$. The Fourier transform of
propagator $\hat{T}(\pm t)$ is $\hat{T}(\pm \omega)=\left(\mp
i\omega\hat{I}-\bf{M}\right)^{-1}\ $, where $\hat{I}$ is an unit
matrix. We can further simplify this expression by using the
spectral decomposition of the matrix $\bf{M}$:
\bq
\mathbf{M} = \sum_{n} \lambda_{n} \mathbf{S} \mathbf{E}^{(nn)} \mathbf{S}^{-1} = 
\sum_{\lambda} \lambda \hat{P}_{\lambda},
\eq
where $\lambda$ are eigenvalues of the matrix
$\bf{M}$, $\mathbf{S}$ is a matrix whose columns are eigenvectors
of $\bf{M}$, $\mathbf{E}^{(nn)}$ is a matrix that has $1$ at $nn$
place and all other elements are zeros, and $\hat{P}_{\lambda}$ is
a projector operator associated with the eigenvalue $\lambda$, so that $\hat{T}(\pm 
\omega)$ is
\bq 
\hat{T}(\pm \omega) = \sum_{\lambda} \frac{\hat{P}_{\lambda}}{\mp
i\omega-\lambda}. 
\eq
Inserting expression for propagator
$\hat{T}$ into Eq.~(\ref{current correlation}) current-current
correlation in the $\omega$-space becomes
\bn
\label{cc}
\langle I_{\eta}(t) I_{\eta'}(0) \rangle_{\omega} &=& \sum_{\lambda,k}
\left [ \frac{\hat{\Gamma}_{\eta}\hat{P}_{\lambda}\hat{\Gamma}_{\eta'}} 
{{-i\omega-\lambda}} {\bm \rho}^{(0)} \right ]_{k} \cr
&+& \sum_{\lambda,k} \left [ 
\frac{\hat{\Gamma}_{\eta'}\hat{P}_{\lambda}\hat{\Gamma}_{\eta}} {{i\omega-\lambda}} {\bm 
\rho}^{(0)} \right ]_{k}.
\en
Eventually, substituting Eq.~(\ref{cc}) into the noise definition 
Eq.~(\ref{powerspectrum}), and noting that summation over zero eigenvalue will be 
canceled out exactly by the term $\langle I_{\eta} \rangle \langle I_{\eta'} \rangle$, 
we can obtain the final expression for a noise power
spectrum: 
\bn
\label{noise} 
&&
S_{I_{\eta}I_{\eta'}}(\omega)=\delta_{\eta\eta'}S^{\rm Sch}_{\eta}\nonumber\\
&& + 2\sum_{k,\lambda \neq 0} \left(
\frac{\left [ \hat{\Gamma}_{\eta} \hat{P}_{\lambda}\hat{\Gamma}_{\eta'} {\bm \rho}^{(0)} 
\right ]_{k}}{{-i\omega-\lambda}} + \frac{ \left [ \hat{\Gamma}_{\eta'} 
\hat{P}_{\lambda} \hat{\Gamma}_{\eta} {\bm \rho}^{(0)} \right ]_{k}} {{i\omega-\lambda}}
\right), \nonumber \\
\en
where $S^{\rm Sch}_{\eta}$ is the frequency-independent Schottky noise originated from 
the self-correlation of a given tunneling event with itself, which the double-time 
correlation function Eq.~(\ref{current correlation}) can not contain. Due to the fact 
that the current has no explicit dependence on the nondiagonal elements of the density 
matrix, it can be simply written as:\cite{Hershfield}    
\bn
\label{Schottky}
S^{\rm Sch}_{\eta} = \sum_{k} \left| \left [ \hat{\Gamma}_{\eta} {\bm \rho}^{(0)} \right 
]_{k}\right|.
\en
It can be shown that in the
zero-frequency limit, $\omega=0$, $S(0)=S_{I_L I_L}=S_{I_R I_R}=-S_{I_L I_R}=-S_{I_R 
I_L}$.
The Fano factor, which
measures a deviation from the uncorrelated Poissonian noise, is
defined as:
\bn
F=\frac{S(0)}{2eI},
\label{Fano factor}
\en
where $2eI$ is the Poissonian noise.

\section{Numerical calculations and discussions}

In the following we consider two magnetic configurations: the
parallel (P), when the majority of electrons in both leads point
in the same direction, chosen to be the electron spin-up state,
$\sigma=\uparrow$;
and the antiparallel (AP), in which the magnetization of the right electrode is 
reversed. The ferromagnetism of the leads can be accounted for by
means of polarization-dependent coupling constants. Thus, we set
for P alignment 
\bq
\Gamma_{L\uparrow}=\Gamma_{R\uparrow}=(1+p)\Gamma, \, 
\Gamma_{L\downarrow}=\Gamma_{R\downarrow}=(1-p)\Gamma, 
\eq
while for AP configuration we choose
\bq
\Gamma_{L\uparrow}=\Gamma_{R\downarrow}=(1+p)\Gamma, 
\,\Gamma_{L\downarrow}=\Gamma_{R\uparrow}=(1-p)\Gamma. 
\eq
Here, $\Gamma$ denotes the tunneling coupling between the QD and the leads without any 
internal magnetization, whereas $p$ ($0\leq p< 1$) stands for the polarization strength 
of the leads. We work in the wide band limit, i.e. $\Gamma$ is supposed
to be a constant, and we use it as an energy unit in the rest of
this paper.
The zero of energy is chosen to be the Fermi level of the left and the right leads in 
the equilibrium condition ($\mu_{L}=\mu_{R}=0$). For clarity, the
bias voltage, $V$, between the source and the drain is considered
to be applied symmetrically, $\mu_{L}=-\mu_{R}=eV/2$. The shift of
the discrete level due to the external bias is neglected.

As a reference case for our analysis we use the analytic result
for the case of paramagnetic electrodes, $p=0$. This case is
exactly solvable even for different couplings to the left and the
right lead, $\Gamma_{L}$ and $\Gamma_{R}$. The resulting Fano
factor is spin-flip independent. In Coulomb blockade regime,
$eV/2<\epsilon_{d}+U$, it is given by
\bq
F=1-\frac{4\Gamma_{L}\Gamma_{R}}{(\Gamma_{L}+2\Gamma_{R})^2}.
\label{fano1.1},
\eq
where the bias voltage is considered to be large
($eV/2\gg\epsilon_{d}$), so that $\Gamma_{L}^{-}=\Gamma_{R}^{+}=0$
and $\Gamma_{L}^{+}=\Gamma_{L}$, $\Gamma_{R}^{-}=\Gamma_{R}$. The
Fano factor depends only on the asymmetry in the coupling between
the leads and the dot: it is equal to $\frac{5}{9}$ for the
completely symmetric case $\Gamma_{L}=\Gamma_{R}$, and approaches
to $1$ when one of the coupling constants becomes much larger than
the other one.

In the opposite regime, when the energy $\epsilon_{d}+U$ is far
below the Fermi level $\mu_{L}$ ($eV/2\gg \epsilon_{d}+U$), we
have $\Gamma_{L}^{-}=\Gamma_{R}^{+}=0$,
  $\tilde{\Gamma}_{L}^{-}=\tilde{\Gamma}_{R}^{+}=0$ and
  $\tilde{\Gamma}_{L}^{+}=\Gamma_{L}^{+}=\Gamma_{L}$,
  $\tilde{\Gamma}_{R}^{-}=\Gamma_{R}^{-}=\Gamma_{R}$, and the Fano
factor is
\bq
F=1-\frac{\Gamma_{L}\Gamma_{R}}{(\Gamma_{L}+\Gamma_{R})^2}. \label{fano2}
\eq
It is equal to $\frac{1}{2}$ for completely symmetric couplings
and to $1$ for the asymmetric ones.

Results of our numerical calculations for the current-voltage
characteristic and the dependence on the Fano factor vs. bias
voltage for P and AP configuration are presented in Figs.~1-3. In
the following we set $\epsilon_{d}=1$, Coulomb interaction $U=4$.
By applying the bias voltage we are varying the fermi levels of
the leads. Two steps in the current voltage characteristic occur:
one is when the Fermi level of the source, $\mu_{L}$, crosses the
discrete levels $\epsilon_{d}$ (for $eV/2>\epsilon_{d}$) and the
other is when the Fermi level $\mu_{L}$ crosses $\epsilon_{d}+U$
(for $eV/2>\epsilon_{d}+U$).

The effects of the polarization on the Fano factor without the
spin-flip scattering ($R=0$) are plotted in Fig.~1. An increase of
the polarization will lead to an enhancement of the current noise
in both configurations (P and AP) but for different reasons. Let us
consider P and AP configurations separately.

\begin{figure}[htb]
\includegraphics[height=8.cm,width=8.5cm]{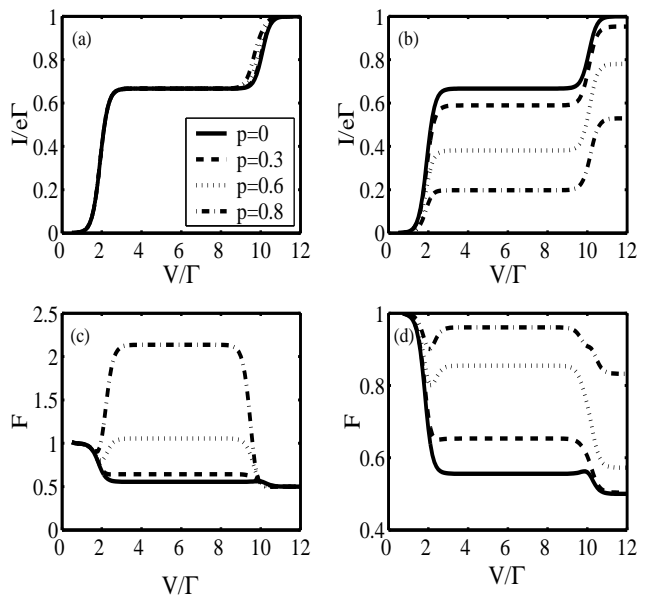}
\caption{Current (a, b) and Fano factor (c, d) vs. the bias voltage
in the P configuration (a, c) and AP configuration (b, d)
calculated for no spin-flip processes and different polarization
$p$. Other parameters are: $\epsilon_{d}=1$, $U=4$,
$T=0.1$.}\label{FIG.1}
\end{figure}

When the leads are in a P configuration [Fig.~1(a),(c)], an
increase of the polarization increases the tunneling rates
$\Gamma_{L\uparrow}$ and $\Gamma_{R\uparrow}$ for electrons with
the spin-up and decreases the tunneling rates
$\Gamma_{L\downarrow}$ and $\Gamma_{R\downarrow}$ for spin-down
electrons. This will increase the spin-up current and decrease the
spin-down current but it will not affect the total current through
the system, which is equal to the sum of the spin-up and spin-down
current. In the limit where the Coulomb interaction prevents a
double occupancy of the dot, there will be competition between
tunneling processes for electrons with the spin-up and those with
the spin-down. The characteristic time for these two processes,
due to polarization, is unequal: there is fast tunneling of
spin-up electrons and slow tunneling of spin-down electrons
through the system. The spin which tunnel with a lower rate
modulate tunneling of the other spin-direction (so-called
dynamical spin-blockade).\cite{Cottet1,Cottet2,Cottet3} Eventually, for a large
value of polarization, it leads to an effective bunching of
tunneling events and, consequently, to the supper-Poissonian shot
noise.

Increasing the bias voltage above the Coulomb blockade regime,
i.e, for $eV/2>\epsilon_{d}+U$, opens one more conducting channel
and removes spin-blockade. In this regime, spin-up and spin-down
electrons are tunneling through the different channels and there
is no more competition between these two tunneling events. This
leads to a reduction of the current fluctuation and the Fano
factor becomes the same as in the paramagnetic case.

The situation is completely different in the AP configuration
[Fig.~1(b),(d)]. An increase of the polarization increases
tunneling rates $\Gamma_{L\uparrow}$ and $\Gamma_{R\downarrow}$
and decreases tunneling rates $\Gamma_{L\downarrow}$ and
$\Gamma_{R\uparrow}$. An electron with the spin-up, which has
tunneled from the left electrode into the QD, remains there for a
long time because the tunneling rate $\Gamma_{R\uparrow}$ is
reduced by the polarization. This decreases the spin-up current.
An increase of the polarization also decreases the spin-down
current because it reduces the probability for
tunneling of the spin-down electrons into the QD. This will decrease
a total current through the system. The enhancement of the noise
in the AP configuration is due to the asymmetry in the tunneling
rates into and out of the QD
($\Gamma_{L\uparrow}>\Gamma_{R\uparrow}$ but
$\Gamma_{L\downarrow}<\Gamma_{R\downarrow}$) for each spin
separately.

\begin{figure}[htb]
\includegraphics[height=8.cm,width=8.5cm]{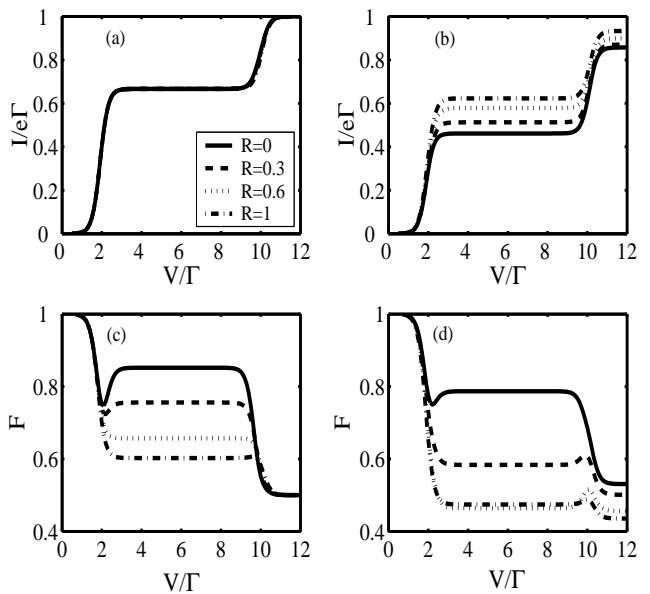}
\caption{Current (a, b) and Fano factor (c, d) in the P
configuration (a, c) and AP configuration (b, d) calculated for
polarization $p=0.5$ and different spin-flip processes. Other
parameters are the same as in Fig.~\ref{FIG.1}.}\label{FIG.2}
\end{figure}

For large voltage, in the regime $eV/2>\epsilon_{d}+U$, both
conducting channels become available which results in reduction of
the noise comparing with the Coulomb blockade regime. In this case
the Fano factor does not go to the paramagnetic value because the
asymmetry in the tunneling rates are still present.

\begin{figure}[htb]
\includegraphics[height=8.cm,width=8.5cm]{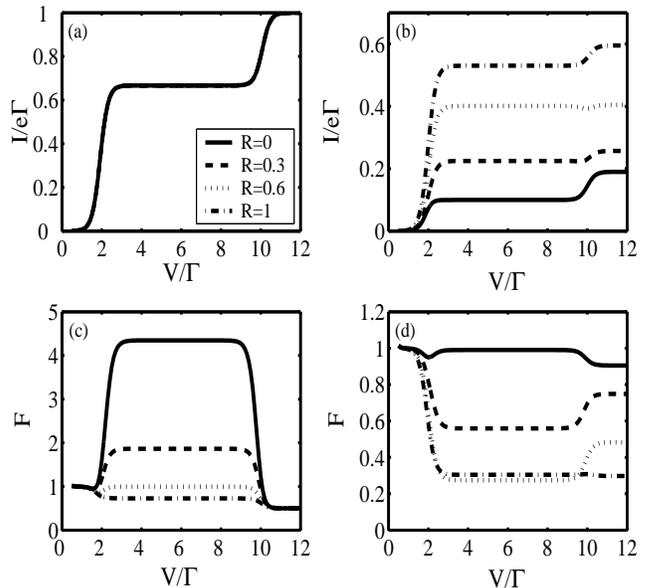}
\caption{The same as Fig.~\ref{FIG.2}\ except for $p=0.9$.}\label{FIG.3}
\end{figure}

The Fano factor dependence on spin-flip scattering can be analyzed
from Figs.~2 and 3. The spin-flip scattering will open one more
path in the tunneling of an electron out of the QD: an electron
with a spin-up(down) can tunnel into the QD, and due to the
spin-flip scattering, it can tunnel out of the QD as an electron
with spin-down(up). An electron which was spending more time in QD
due to polarization (spin-down in the P configuration and spin-up
in the AP configuration) now has more probability to tunnel out
from the dot. This causes the decrease of the current fluctuation
and Fano factor. However, this is not true for the P-configuration
when double occupation of the QD is allowed (for
$eV/2>\epsilon_{d}+U$). In this regime the spin-flip does not have
any effect on the Fano factor [Figs.~2(c) and 3(c)]. Spin-up and
spin-down electrons are passing through separate channels without
changing their spins.

\section{CONCLUSION}

In this paper we have analyzed the zero frequency current shot
noise through a quantum dot connected to two ferromagnetic leads, taking account
for the effects caused by the Coulomb interaction and weak
intradot spin-flip scattering. For this purpose the rate equation
approach for the description of the shot noise was modified to
include the coherent electron evolution inside of the dot.
We calculated the Fano factor
dependence on the lead configuration (parallel and antiparallel configurations),
the degree of their polarization and the amplitude of the
spin-flip process. Our results readily show the regimes when the
information provided by shot noise measurement will significantly
exceed the information which can be obtained from measuring the
current.

\section{Acknowledgement}

This work was supported by the DURINT Program administered by the US Army Research 
Office. One of the authors, I. Djuric, acknowledges valuable discussions with L. 
Fedichkin and V. Puller.


\begin{thebibliography}{99}

\bibitem[$\dagger$]{db}{Corresponding author e-mail: bdong@stevens.edu}

\bibitem{Prinz}{G.A. Prinz, ``Magnetoelectronics," \emph{Science}, vol. 282, pp. 
1660-1663, 1998.}

\bibitem{Wolf}{S.A. Wolf, D.D. Awschalom, R.A. Buhrman, J.M. Da ughton, S.von Molnar, 
M.L. Roukes, A.Y. Chtchelkanova, and D.M. Treger, ``Spintronics: A Spin-Based 
Electronics Vision for the Future," \emph{Sience}, vol. 294, pp. 1488-1495, 2001.}

\bibitem{dot1}{J. Barna\`{s} and A. Fert, ``Magnetoresistance Oscillations due to 
Charging Effects in Double Ferromagnetic Tunnel Junctions," \emph{Phys. Rev. Lett.}, 
vol. 80, pp. 1058-1061, 1998.}

\bibitem{dot2}{S. Takahashi and S. Maekawa, ``Effect of Coulomb
Blockade on Magnetoresistance in Ferromagnetic Tunnel Junctions,"
\emph{Phys. Rev. Lett.}, vol. 80, pp. 1758-1761, 1998.}

\bibitem{dot3}{X.H. Wang and A. Brataas, ``Large Magnetoresistance Ratio in 
Ferromagnetic Single-Electron Transistors in the Strong Tunneling Regime," \emph{Phys. 
Rev. Lett.}, vol. 83, pp. 5138-5141, 1999.}

\bibitem{Fedichkin}{D. Mozyrsky, L. Fedichkin, S.A. Gurvitz, and G.P. Berman, 
``Interference effects in resonant magnetotransport," \emph{Phys.
Rev. B}, vol. 66, p. 161313, 2002.}

\bibitem{Erlingsson}{S.I. Erlingsson, and Yu.V. Nazarov, ``Hyperfine-mediated 
transitions between a Zeeman split doublet in GaAs quantum dots: The role of the 
internal field," \emph{Phys. Rev. B} vol. 66, p. 155327, 2002.}

\bibitem{Beenakker}{C. Beenakker and C. Sch\"onenberger, ``Quantum shot noise," 
\emph{Phys. Today}, vol. 56 (5), pp. 37-42, May 2003.}

\bibitem{Blanter}{For an overview of quantum shot noise, please refer to Ya.M. Blanter 
and M. B\"uttiker, ``Shot Noise in Mesoscopic Conductors,"
\emph{Phys. Rep.}, vol. 336, p. 1, 2000.}

\bibitem{Hershfield}{S. Hershfield, J.D. Davies, P. Hyldgaard, C.J. Stanton and J.W. 
Wilkins, ``Zero-frequency current noise for the double-tunnel-junction Coulomb 
blockade," \emph{Phys. Rev. B}, vol. 47, pp. 1967-1979, 1993.}

\bibitem{Korotkov}{A.N. Korotkov, ``Intrinsic noise of the single-electron transistor," 
\emph{Phys.Rev. B}, vol. 49, pp. 10381-10392, 1994.}

\bibitem{Hanke}{U. Hanke, Yu.M. Galperin, K.A. Chao and Nanzhi Zou, ``Finite-frequency 
shot noise in a correlated tunneling current," \emph{Phys. Rev. B}, vol. 48 pp. 
17209-17216, 1993}

\bibitem{Nazarov}{Yu.V. Nazarov, ``Quantum Interference, Tunnel Junctions and Resonant 
Tunneling," \emph{Physica B}, vol. 189, pp. 57-69, 1993.}

\bibitem{Gurvitz}{S.A. Gurvitz and Ya.S. Prager, ``Microscopic derivation of rate 
equations for quantum transport," \emph{Phys. Rev. B}, vol. 53, pp. 15932-15943, 1996.}

\bibitem{Dong}{Bing Dong, H.L. Cui, and X.L. Lei, ``Quantum rate equations for electron 
transport through an interacting system in the sequential tunneling regime,"
\emph{Phys. Rev. B}, vol. 69, pp. 35324-035339, 2004.}

\bibitem{Cottet1} {A. Cottet and W. Belzig, ``Dynamical spin-blockade in a quantum dot 
with paramagnetic leads," \emph{Europhys. Lett.}, vol. 66, pp. 405-411, 2004.}

\bibitem{Cottet2}{A. Cottet, W. Belzig, and C. Bruder, ``Positive Cross Correlations in 
a Three-Terminal Quantum Dot with Ferromagnetic Contacts," \emph{Phys. Rev. Lett.}, vol. 
92, pp. 206801, 2004.}

\bibitem{Cottet3}{A. Cottet, W. Belzig, and C. Bruder, ``Positive cross-correlations due 
to Dynamical Channel-Blockade in a three-terminal quantum dot," \emph{cond-mat/0403507}, 
2004.}

\end{thebibliography}
\end{document}